# A Comparative Experimental and Theoretical Study on Doubly Differential Electron-Impact Ionization Cross Sections of Pyrimidine

M. Dinger[1,2,*], W. Y. Baek[1], and H. Rabus[3]

[1]Physikalisch-Technische Bundesanstalt (PTB), Bundesallee 100, 38116 Braunschweig, Germany
[2]Ruprecht-Karls-Universität Heidelberg, Grabengasse 1, 69117 Heidelberg, Germany
[3]Physikalisch-Technische Bundesanstalt (PTB), Abbestrasse 2-12, 10587 Berlin, Germany

[*]Corresponding author: mareike.dinger@ptb.de

**Abstract**

To provide a comprehensive data set for track structure-based simulations of radiation damage in DNA, doubly differential electron-impact ionization cross sections of pyrimidine, a building block of the nucleobases cytosine and thymine, were measured for primary electron energies between 30 eV and 1 keV as a function of emission angle and secondary electron energy. The measurements were performed for secondary electron energies from 4 eV to about half of the primary electron energy and for emission angles between 25° and 135°. Based on the experimental doubly differential ionization cross sections, singly differential and total ionization cross sections of pyrimidine were determined and compared to calculations using the BEB model. In addition to the measurements, a theoretical approach for calculating triply and doubly differential ionization cross section was developed, which is based on the distorted wave Born approximation, a single center expansion of molecular orbitals and an averaging of the **T**-matrix over different molecular orientations. The calculated doubly differential ionization cross sections of pyrimidine show a qualitatively good agreement with the experimental results.

**I. Introduction**

It is well-established that DNA damage is the primary mechanism associated with radiation-induced carcinogenesis [1]. Ionizing radiation can induce cellular damage either through direct excitation and ionization of DNA or via secondary electrons. The latter are produced usually in a large number during the penetration of ionizing radiation in tissue. The contribution of secondary electrons to radiation damage becomes particularly significant in the case of low linear energy transfer (LET) radiations.

The biological impact of ionizing radiation is influenced not only by the amount of energy deposited but also by its track structure at a sub-micrometer scale [2]. Radiation tracks, and consequently the biological effectiveness, vary significantly with radiation type, and understanding this variation is important for treatment optimization in radiotherapy, especially in modern treatment modalities using protons and carbon ions. The biological effectiveness of radiation in human tissue has been often assessed through track structure simulations in water [3-5]. It is assumed here that human tissue can be wholly represented by water of different densities. However, several studies indicate that the electron transport property of water is significantly different from that of other biological media. For instance, electron kinetic calculations by White et al. [6] showed distinct differences between the transport coefficients of electrons in $H_2O$ and tetrahydrofuran, a surrogate for deoxyribose. In addition, it was also shown that track structure simulations based on water only underestimate the damage induced by radiation in DNA [7].

For a more realistic modeling of radiation track structure in human tissue, significant efforts have been made to obtain comprehensive data sets for electron interaction cross sections of biomolecules. A particular focus has been placed on the electron interaction cross sections of the molecular components of DNA. One of these components is pyrimidine (Py) which is the core building block of the nucleobases cytosine and thymine. Fuss et al. [8] reported the total electron scattering cross section of Py in the energy range between 8 eV and 500 eV. Maljković et al. [9] published differential elastic scattering cross sections of Py for electron energies from 50 eV to 300 eV. Similar measurements were performed by Palihawadana et al. [10], who focused on lower electron energies ranging from 3 eV to 50 eV. Regarding the electron-impact ionization of Py, Linert et al. [11] obtained the total ionization cross section (TICS) of Py by collecting its ionic fragments for electron energies from the ionization threshold to 150 eV. Builth-Williams et al. [12] reported triply differential ionization cross sections (TDCS) of Py for selected secondary and primary electron energies of 20 eV and 250 eV, respectively, at a few scattering angles. They also calculated the TDCS of Py using the molecular 3-body distorted wave approximation.

Following the measurement of total [13] and differential elastic electron scattering cross sections [14], we determined the doubly differential ionization cross section (DDCS) of Py experimentally as well as theoretically as a function of emission angle $\theta$ and secondary electron energy $E$ for primary electron energies $T$ from 30 eV to 1 keV. The measured range of emission angles and secondary electron energies spanned from 25° to 135° and from 4 eV to ($T$-$I$)/2, respectively, where $I$ is the ionization threshold of Py. The DDCS beyond the measured angular range was obtained by extrapolating the experimental DDCS with the help of a semi-empirical model. In addition to the measurements, the DDCS of Py was calculated using a theoretical approach that was derived based on the distorted wave Born approximation (DWBA) and the single center expansion [15] of multi-centered molecular orbitals. In the calculation, the DDCS was obtained by integrating the triply differential ionization cross sections (TDCS) that were properly averaged over different molecular orientations. This differs from many existing approaches [12,16-18], where molecular orbitals (MO), not the TDCS itself, were averaged over various orientations. As pointed out by Gao et al. [19], the orientation averaging of MO (OAMO) is only suitable in the case when the MO is dominated by s-basis functions.

## II. Experimental method

The DDCS of Py was determined absolutely using an experimental method that has been explained in detail in our earlier works [14,20,21]. Therefore, a brief recapitulation of the experimental principle is given below. Figure 1 shows a schematic view of the experimental setup. The DDCS was measured using a crossed-beam apparatus where an electron beam of primary energy $T$ perpendicularly intersects an effusive molecular beam. The current $\bar{I}$ of the electron beam was measured with a Faraday cup placed beyond the molecular beam. Electrons emitted from the interaction zone were analyzed according to their energy $E$ using a hemispherical deflection analyzer. The emission angle $\theta$ of electrons to be detected in the scattering plane is given by the angle between the electron beam direction and the central view axis of the hemispherical deflection analyzer. It could be changed by rotating the electron gun that was mounted on a turntable.

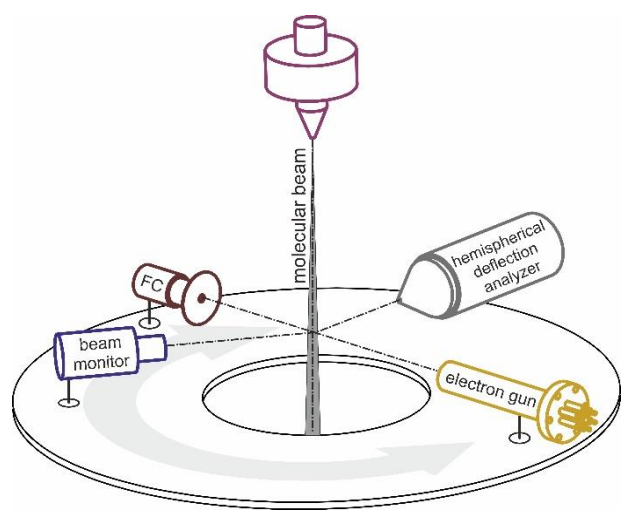

Fig. 1. Schematic view of the measurement arrangement, comprising several key components: an electron gun, a Faraday cup (FC), a beam monitor, an effusive molecular beam, and a hemispherical deflection analyzer. The electron gun, FC and beam monitor were affixed directly to a turntable, while the capillary for the molecular beam could be concurrently rotated alongside the turntable, both revolving around the same axis. The experimental setup was housed within a scattering chamber constructed from 8 mm thick permalloy to shield against the Earth's magnetic field.

The count rate $\Delta\dot{N}/\Delta E\Delta\Omega$ per energy interval $\Delta E$ and solid angle $\Delta\Omega$ is related to the DDCS $d^2\sigma/dEd\Omega$ via

$$\frac{1}{\eta(E)}\frac{\Delta\dot{N}(\theta)}{\Delta E\Delta\Omega} = \frac{-\bar{I}_0}{e}\tilde{V}_{\text{eff}}\frac{d^2\sigma}{dEd\Omega}(\theta, E), \quad (1)$$

where $\eta(E)$ is the energy-dependent electron detection efficiency, $\bar{I}_0$ is the primary electron beam current, $e$ is the elementary charge, and $\tilde{V}_{\text{eff}}$ is the effective number of interacting molecules per area. For the calculation of $\tilde{V}_{\text{eff}}$, the spatial distribution of electrons and molecules in the interaction zone must be known [22,23]. As this calculation is hardly feasible in practice, measurements of DDCS using a molecular beam are commonly conducted by applying the relative flow technique [24]. Instead of using the relative flow technique, $\tilde{V}_{\text{eff}}$ was assessed in the present work from the attenuation of the primary electron beam current after it passed the molecular beam:

$$\frac{\Delta\bar{I}}{e} = \frac{|\bar{I}_0 - \bar{I}|}{e} = \frac{-\bar{I}_0}{e}\tilde{V}_{\text{eff}} \times \sigma_t, \quad (2)$$

where $\sigma_t$ is the total electron scattering cross section of the molecule of interest. Equation (2) simply states that the total rate of electrons scattered out of the primary electron beam in the interaction zone is equal to the loss of the number of primary electrons per second in the forward direction.

For the determination of the detection efficiency $\eta(E)$ its relative energy dependence $\eta_r(E)$, defined by $\eta_r(E) = \eta(E)/\eta(20\text{eV})$ was first calculated by simulating electron transport in phase space through the hemispherical deflection analyzer. It was here made use of the fact that the calculation of the analyzer transmission on a relative scale is much more accurate than on an absolute scale. The absolute detection efficiency $\eta(E)$ was then obtained by multiplying the computed $\eta_r(E)$ with the measured detection efficiency $\eta_{\text{ex}}$ at 20 eV:

$$\eta(E) = \eta_r(E) \times \eta_{\text{ex}}(20\text{eV}). \quad (3)$$

The efficiency $\eta_{\text{ex}}(20\text{eV})$ was determined using elastic scattering of 20 eV- electrons by helium. As the inelastic scattering cross sections of He are negligibly small for $T \leq$ 20 eV, the current loss of a 20 eV-electron beam in He occurs almost entirely due to elastic scattering. Therefore, $\eta_{\text{ex}}(20\text{eV})$ can be obtained from the measured integral elastic count rate $\dot{N}_{\text{el}}$ and the current loss $\Delta I$ via the relation

$$\dot{N}_{\text{el}} = \int \frac{\Delta\dot{N}_{\text{el}}}{\Delta\Omega}d\Omega = \eta_{\text{ex}}(20\text{eV}) \times \frac{\Delta I}{e}. \quad (4)$$

The determination of the detection efficiency was explained in detail in our earlier publications [20,21].

**III. Measurement**

The molecular target was produced by the effusion of Py vapor through a cylindrical gas nozzle 2 mm in diameter and 80 mm in length. The purity of Py stated by the supplier, Aldrich Chemical Ltd., was better than 99%. The molecular flow rate was adjusted by regulating the driving pressure above the gas nozzle, typically set at 0.5 mbar. This pressure resulted in an effective molecular area density on the order of $5 \times 10^{13}$ cm$^{-2}$, which was low enough to fulfil the single collision condition, but sufficiently high to cause attenuation of the primary electron beam by more than 3%. When varying the electron detection angle, the gas nozzle was rotated synchronously with the electron gun around the same axis to avoid changes in $\tilde{V}_{\text{eff}}$ with $\theta$. Temporal fluctuations of $\tilde{V}_{\text{eff}}$ were monitored by counting electrons emitted from the interaction zone with a channel electron multiplier (beam monitor) that was placed at a fixed angular position relative to the electron gun.

The molecular beam was crossed by the electron beam 1 mm below the gas nozzle. Depending on the electron energy $T$, the primary beam current varied between 10 pA and 1 nA. The beam current was raised at higher electron energies to compensate for the decrease in DDCS with increasing $T$. An upper limit was set on the electron beam current to ensure that the count rate did not exceed $10^4$ s$^{-1}$, as exceeding this range resulted in a noticeable decrease in detection efficiency. The energy width (full width at half maximum) of the electron beam was less than 0.5 eV. It is noteworthy that the experiment was conducted in a scattering chamber made of permalloy of 8 mm in thickness. The residual magnetic field in the scattering plane was lower than 1 μT.

As mentioned above, a hemispherical deflection analyzer was used to measure the energy spectra of electrons originating from the interaction zone. The analyzer, with a mean radius of 150 mm and a deflection angle of 180°, was equipped with an array of 5 channel electron multipliers for electron detection. The angular resolution as well as the acceptance of the analyzer could be adjusted using an iris aperture located at its entrance. For emission angles above 35°, the half acceptance angle of the analyzer was set to 1.5°, whereas it was adjusted to 0.8° for $\theta < 35°$. The analyzer was operated in the Contant-Retard-Ratio mode [25], in which the ratio of the initial kinetic energy of electrons to the pass energy was kept constant. Depending on the energy range, different retard ratios were employed to achieve the energy resolution between 1.7 eV at $T$=1 keV and better than 0.5 eV for $T \leq 100$ eV. The angle-positioning accuracy of the electron gun was examined using the differential elastic scattering cross section of Ar, which exhibits a resonance-like structure at electron energies around 100 eV. It was found that the positioning of the electron gun could be reproduced within 2°.

Gaseous molecules effusing through the gas nozzle led to an increase in the residual pressure within the scattering chamber. Since incident electrons could also be scattered by the residual gas, a background spectrum was measured for each detection angle and subsequently subtracted from the main spectrum obtained using the molecular beam. To measure the background spectrum, Py vapor was diffusely introduced not through the gas nozzle but through a separate valve connected to a wide aperture situated on the wall of the scattering chamber. The gas flow rate through the valve was adjusted such that the pressure in the scattering chamber was equal to the residual pressure during the measurement of the main spectrum. The ratio of the count rates in the background measurement to that in the main spectrum measured with the molecular beam was approximately 4%.

**IV. Theoretical method**

The DDCS as a function of the solid angle $\Omega_B$ and energy $E_B$ of the ejected electrons was obtained by integrating the triply differential ionization cross section (TDCS) over the solid angle $\Omega_A$ of the scattered electrons and summing over all target orbitals. For a closed-shell molecule undergoing electron impact with initial momentum $k_0$, the TDCS for ejection of an electron from the $i$-th molecular orbital (MO) in atomic units can be expressed as [26]:

$$\frac{d^3\sigma_i}{d\Omega_A d\Omega_B dE_B} = (2\pi)^4 \frac{k_A k_B}{k_0} \frac{n_i}{4} \sum_S (2S+1)\left|T_S(\boldsymbol{k_0},\boldsymbol{k_A},\boldsymbol{k_B})\right|^2, \tag{5}$$

where $k_A$ and $k_B$ are the momenta of the scattered projectile and ejected electron, respectively, $n_i$ is the number of electrons in the $i$-th molecular orbital (MO), and $T_S$ is the **T**-matrix for the two-electron spin $S$ given by the spins of the projectile and the collision partner. $S$ can take the values 0 (singlet) and 1 (triplet). Starting from the DWBA formalism for atoms given by McCarthy and Xixiang [26], we derive a theoretical approach for the orientation-averaged DDCS of polyatomic molecules. The **T**-matrix for the spin state $S$ can be expressed by

$$T_S(\boldsymbol{k_0},\boldsymbol{k_A},\boldsymbol{k_B}) = \langle \chi^-(\boldsymbol{k_A},\boldsymbol{r_A})\chi^-(\boldsymbol{k_B},\boldsymbol{r_B})|V_{ee}[1+(-1)^S P_r]|\psi_i(\boldsymbol{r_B})\chi^+(\boldsymbol{k_0},\boldsymbol{r_A})\rangle, \tag{6}$$

where the final state is represented by the product of the distorted waves $\chi^-$ of the scattered projectile and ejected electron. The initial state is given by the target electron wave function $\psi_i(\boldsymbol{r_B})$ times the distorted wave $\chi^+$ of the incident projectile electron. $V_{ee}$ is the interaction potential between the projectile and target electron and $P_r$ is the space exchange operator [26]. The combination of Eqs. (5) and (6) results in

$$\frac{d^3\sigma_i}{d\Omega_A d\Omega_B dE_B} = (2\pi)^4 \frac{n_i k_A k_B}{k_0}[|T_{\text{dir}}|^2 + |T_{\text{ex}}|^2 - Re(T_{\text{dir}}^* T_{\text{ex}})] \tag{7}$$

with

$$T_{\text{dir,ex}}(\boldsymbol{k_0},\boldsymbol{k_A},\boldsymbol{k_B}) = \langle \chi^-(\boldsymbol{k_A},\boldsymbol{r_{A,B}})\chi^-(\boldsymbol{k_B},\boldsymbol{r_{B,A}})|V_{ee}|\psi_i(\boldsymbol{r_B})\chi^+(\boldsymbol{k_0},\boldsymbol{r_A})\rangle, \tag{8}$$

The initial distorted wave function $\chi^+$ of the projectile with the kinetic energy $E_0$ is the solution of the Schrödinger equation for elastic scattering

$$(K_0 + U_i)\chi^+ = E_0\chi^+ \tag{9}$$

and the final distorted wave functions $\chi^-_{A,B}$ are the solutions of the Schrödinger equation

$$(K_{A,B} + U_f)\chi^-_{A,B} = E_{A,B}\chi^-_{A,B}, \tag{10}$$

where $K_0$, $K_A$, and $K_B$ are the respective kinetic energy operators. In the DWBA, the spherically symmetric optical potential of the molecule is usually employed as the initial distorting potential $U_i$ for the incident electron wave. Similarly, the outgoing electrons, i. e. the scattered projectile and the ejected electron, are distorted by the spherically symmetric electrostatic potential $U_f$ of the residual molecular ion.

The computation of the **T**-matrix can be simplified to a one-dimensional problem by factorizing the wavefunctions into angular and radial parts. The in- and outgoing electron wavefunctions were factorized using the partial wave expansion method:

$$\chi^\pm(\boldsymbol{k},\boldsymbol{r}) = (2\pi)^{-\frac{3}{2}} \left(\frac{4\pi}{kr}\right) \sum_{LM} i^{\pm L} e^{i\sigma_L} \chi_L(k,r) Y_{LM}^*(\hat{\boldsymbol{k}}) Y_{LM}(\hat{\boldsymbol{r}}), \qquad (11)$$

where $Y_{LM}$ is the spherical harmonics function. Likewise, the two-electron Coulomb potential $V_{\text{ee}}$ was factorized using the Laplace expansion:

$$V_{\text{ee}} = \frac{1}{|\boldsymbol{r_A} - \boldsymbol{r_B}|} = \sum_{\lambda\mu} \frac{4\pi}{\hat{\lambda}^2} \frac{r_<^\lambda}{r_>^{\lambda+1}} Y_{\lambda\mu}^*(\hat{\boldsymbol{r}}_A) Y_{\lambda\mu}(\hat{\boldsymbol{r}}_B) \qquad (12)$$

with $\hat{\lambda}^2 = 2\lambda + 1$ and $r_<$, $r_>$ the lesser and greater of $\boldsymbol{r_A}$ and $\boldsymbol{r_B}$.

Unlike atomic orbitals, the wavefunctions contributing to the molecular orbitals are not centered on a single nucleus, making it challenging to factorize them in angular and radial parts. To facilitate this factorization, the multicentered orbital wavefunction $\psi_i(\boldsymbol{r_B}, \boldsymbol{R})$ was expanded around the center of mass using single-center symmetry-adapted angular functions $\mathrm{X}_{hl}^{p_i\mu_i}(\hat{\boldsymbol{r}}_B)$ [27]:

$$\psi_i(\boldsymbol{r_B}, \boldsymbol{R}) = \frac{1}{r_B} \sum_{hl} u_{hl}^i(r_B) \mathrm{X}_{hl}^{p_i\mu_i}(\hat{\boldsymbol{r}}_B). \qquad (13)$$

For simplicity of notation, all coordinates of the nuclei are denoted by $\boldsymbol{R}$ in Eq. (13). The functions $\mathrm{X}_{hl}^{p_i\mu_i}$ possess the property that they transform under the symmetry operations of the molecule's point group in the same way as the irreducible representation $p_i$. The index $\mu_i$ distinguishes different components of the representation, while $h$ distinguishes different bases for given values of $p_i, \mu_i$ and $l$. For molecules like Py belonging to the $C_{2\text{v}}$ point group, where all irreducible representations are one dimensional, the distinction between components of the representations by the index $\mu_i$ is unnecessary. As molecular point groups are subgroups of the full rotational group, and the spherical harmonics constitute a basis of the full rotational group, the symmetry-adapted angular functions $\mathrm{X}_{hl}^{p_i}$ can be expressed as linear combinations of spherical harmonics $Y_{lm}$:

$$\mathrm{X}_{hl}^{p_i}(\hat{\boldsymbol{r}}_B) = \sum_m b_{hlm}^{p_i} Y_{lm}(\hat{\boldsymbol{r}}_B). \qquad (14)$$

For the $C_{2v}$ symmetry group the different irreproducible representations for a given $l$ only refer to one value of $m$ [28]. In the following equations, we will therefore replace the index $h$ in the radial functions $u_{hl}^i(r_B) \equiv u_{lm}^i(r_B)$ with the associated value of $m$. The coefficients $b_{lm}^{p_i}$ can be obtained from the character tables of the irreducible representations of the molecule's point group. In the case of the $C_{2v}$ point group, $p_i$ can take values $A_1, A_2, B_1$, and $B_2$. The radial functions $u_{lm}^i(r_B)$ were computed using the library SCELib4.0 [27]. For conciseness of the main text, further derivations of the computational form of the **T**-matrix are provided in the appendix A1.

The above derivations refer to a coordinate system fixed to the molecule, called molecular frame (MF). It is assumed that the nuclei in the molecule remain fixed, in other words, no molecular vibrations and rotations occur during the ionization process. However, since molecules in the gas beam are randomly oriented, the cross sections defined in the MF must be averaged over different molecular orientations to obtain the cross sections in the laboratory frame (LF). When the same origin is chosen for both the MF and LF, a change in molecular orientation is equivalent to a rotation of the MF relative to the LF. Since the radial parts are invariant against different molecular orientations (see Appendix, Eq. (A2)), only the angular parts of the molecular wavefunctions described by the corresponding spherical harmonics need to be transformed from the MF to the LF. This transformation can be achieved using the Wigner D-matrix $D_{m_1,m}^l(\alpha, \beta, \gamma)$

$$Y_{lm}(\theta, \varphi) = \sum_{m_1=-l}^{l} D_{m_1,m}^l(\alpha, \beta, \gamma) Y_{lm_1}(\tilde{\theta}, \tilde{\varphi}), \tag{15}$$

where $\alpha, \beta$ and $\gamma$ represent the Euler angles, and the angles in the MF are indicated by a tilde. In difference to OAMO, the molecular wavefunction (Eqs. (13)-(14)) is multiplied with the Wigner D-matrix according to Eq. (15) and the absolute square of the resulting **T**-matrix $T'_{\text{dir}}$ is averaged over different orientations

$$\overline{|T_{\text{dir}}|^2} = \frac{1}{8\pi^2} \int_0^{2\pi} d\alpha \int_0^\pi \sin\beta \, d\beta \int_0^{2\pi} d\gamma \, |T'_{\text{dir}}|^2. \tag{16}$$

In principle, the DDCS can be computed by numerically integrating the orientation-averaged TDCS over the solid angle of scattered electrons. To expedite computation, we analytically integrate the orientation-averaged **T**-matrix over $\widehat{\mathbf{k}}_A$:

$$\frac{d^2\sigma_i}{d\Omega_B dE_B} = (2\pi)^4 \frac{n_i k_A k_B}{k_0} \int d\widehat{\mathbf{k}}_A \, \overline{|T_{\text{dir}}|^2} + \text{exchange terms}. \tag{17}$$

The derivation of the computational form of Eq. (17) is provided in appendix A2.

Using Eq. (17) (in numerical form Eqs. (A2, A10-A12)) the DDCS was calculated for each molecular orbital and summed up. The calculation was performed for the 15 valence orbitals, i.e., excluding the four carbon and two nitrogen K-shell orbitals. Molecular wavefunctions were obtained using the Gaussian09 software [29] with the basis set 6-311++G.

**V. Uncertainty analysis**

The uncertainties of the measured DDCS were evaluated following the *Guide to the Expression of Uncertainty in Measurement* [30]. The uncertainty of the effective number of molecules per area $V_{\text{eff}}$, determined from the current loss $\Delta \bar{I}$ of the primary beam current across the molecular beam and from the TCS of Py, was estimated to be 15%. The same uncertainty was attributed to the detection efficiency $\eta$. The uncertainty of the primary beam current $\bar{I}_0$ was determined from the standard deviation of its temporal fluctuations during the measurement, which amounted to 5%. Another source of uncertainty stemmed from the statistical uncertainty of the measured counts of the electron energy spectra, which was at most 10%. Furthermore, the impurity of the Py vapor caused an uncertainty of 2%. As the measured quantities in Eq. (1) are not correlated, the overall relative uncertainty was determined as the square root of the quadratic sum of the individual relative uncertainties, resulting in a value of 24%.

## VI. Results and Discussion

Figure 2 displays the DDCS of Py as a function of the emission angle $\theta$ for various primary and secondary electron energies. The present experimental results are represented by different symbols (depending on secondary electron energy), while the solid curves depict the theoretical values calculated with Eq. (17). The dashed curves are semi-empirical fits to the experimental data, the details of which are elaborated below. The complete experimental results are provided in the supplement material.

It can be seen from Fig. 2 that the experimental DDCS for secondary electron energies around 10 eV exhibit a weak angle dependence, particularly noticeable at high primary electron energies. As the secondary and primary electron energies increase, binary collision peaks become more pronounced. This aligns with the DWBA calculation and with the semi-empirical formula proposed by Rudd [31]. According to Rudd's formula, the width $\Gamma_1$ of the binary collision peak decreases with increasing secondary electron energy $E$:

$$\Gamma_1 = \text{const} \times \left(\frac{1 - \cos^2\theta_0}{E/B}\right)^{1/2}. \tag{18}$$

Here, $B$ is the binding energy of the ejected electron and $\theta_0$ is the position of the binary collision peak, varying with $E$ and $T$:

$$\cos\theta_0 \cong \left(\frac{E+B}{T}\right)^{1/2}. \tag{19}$$

Consistent with Eq. (19), the shift of the binary collision peaks to lower angles with increasing $E$ can be also observed in the present experimental DDCS.

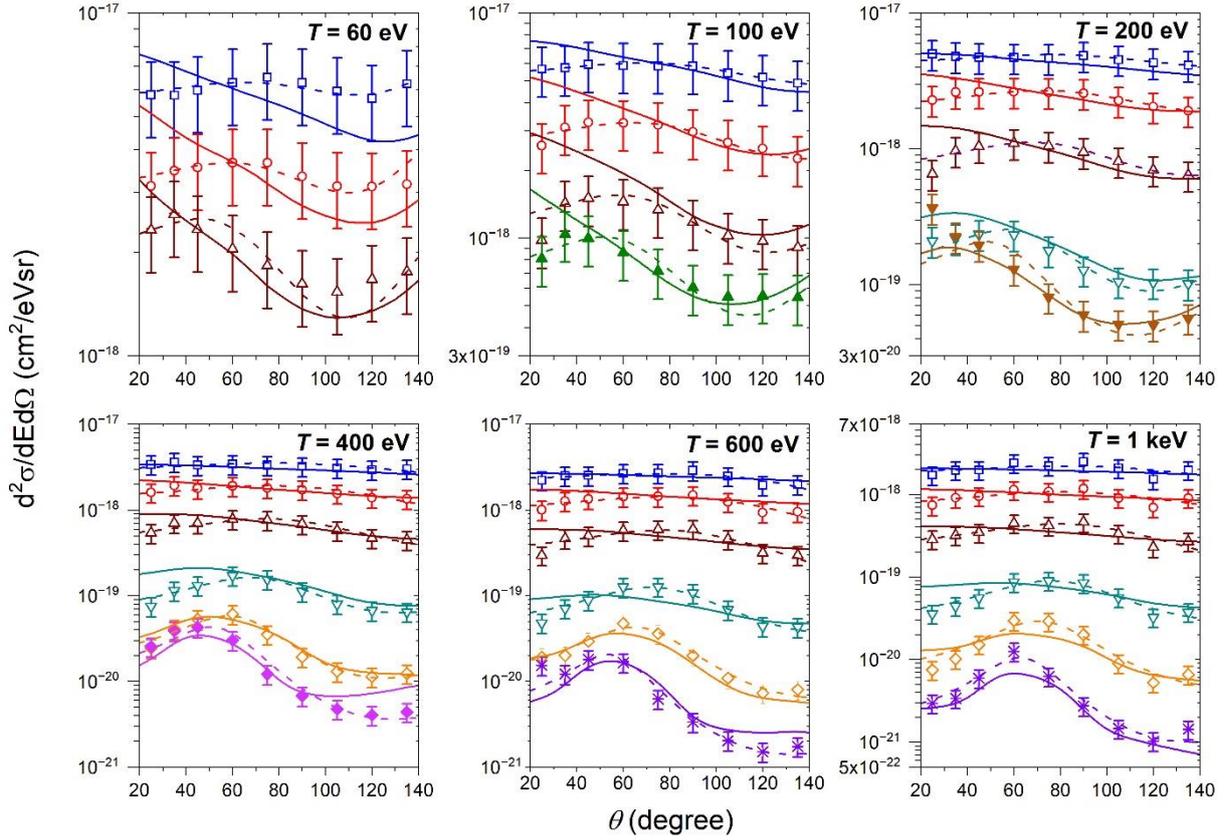

Fig. 2. DDCS of Py as function of emission angle $\theta$ for various primary electron energies $T$. Experimental data points corresponding to different secondary electron energies are indicated by distinct symbols: (□) 5 eV, (○) 10 eV, (△) 20 eV, (▲) 30 eV, (▽) 50 eV, (▼) 80 eV, (◇) 100 eV, (◆) 150 eV, (∗) 200 eV. The solid and dashed curves represent calculations using the DWBA and the best fits with a semi-empirical formula (see below), respectively.

   Overall, the results of the present measurement are satisfactorily reproduced by the DWBA calculation within the experimental uncertainties. However, it's somewhat surprising that the DWBA calculation, typically valid at primary electron energies above a few hundred eV, matches the experimental data as well as it does. Notably, significant differences between both data were observed at emission angles below 30°, where the DWBA calculation tends to overestimate the DDCS. The overestimate appears to be partially caused by the neglect of post collision interaction (PCI) between the scattered and ejected electron in the DWBA approach employed in this study. A preliminary assessment of the influence of PCI suggests that this interaction leads to a reduction of the DDCS at low emission angles. Since including PCI in the analytical formula of DDCS was very difficult, this assessment was made by comparing numerically integrated TDCS with and without the inclusion of PCI according to the formulation of Ward and Macek [32]. In the preliminary evaluation, the integration was carried out with angular increments of 30° due to large computation time required. The reduction increases with decreasing emission angle and becomes significant when the velocities of the scattered and ejected electron are similar. The potential influence of PCI is most evident in the experimental data for $T$=100 eV, where the largest deviation from the DWBA calculation occurs at secondary electron energies above 20 eV. In this energy range, the velocity of the ejected electron approaches that of the scattered projectile electron, which can suffer a kinetic energy loss of up to about 40 eV depending on the binding energy of the ionized MO.

Based on the experimental DDCS presented in Fig. 2, the singly differential ionization cross section (SDCS) $d\sigma/dE$ of Py was determined by integrating the DDCS over the solid angle. The DDCS beyond the measured angular range was obtained by extrapolating the experimental data. For the extrapolation, a semi-empirical formula [31] was fitted to the experimental data. The semi-empirical formula comprises two Lorentzian functions describing the binary collision peak and the angular distribution of electrons ejected in the backward direction, and a constant value denoting the contribution of forward electron emission:

$$\frac{d^2\sigma}{d\varepsilon d\Omega}(\theta,\varepsilon) = a_1[f_{BE}(\theta,\varepsilon) + a_2 f_b(\theta,\varepsilon) + a_3] \qquad (20)$$

with the dimensionless variable $\varepsilon = E/B$. The two Lorentzian functions $f_{BE}$ and $f_b$ are given by

$$f_{BE}(\theta,\varepsilon) = \frac{1}{1 + [(\cos\theta - \cos\theta_0)/\Gamma_1]^2}, \qquad (21)$$

and

$$f_b(\theta,\varepsilon) = \frac{1}{1 + [(\cos\theta + 1)/\Gamma_2]^2}. \qquad (22)$$

The quantities $\Gamma_1$ and $\cos\theta_0$ in Eq. (21) are defined by Eq. (18) and Eq. (19), respectively, while the constants $a_1$, $a_2$, and $a_3$ in Eq. (20) are used as fit parameters. The value of $\Gamma_2$ was fixed to 0.36. It should be noted that the semi-empirical fit should ideally be performed for each MO because the binding energies of non-degenerate MOs vary. Since the contribution of the individual MOs to the experimental DDCS could not be resolved, the fits were carried out by substituting the individual binding energies with the ionization threshold $I$ (9.8 eV) of the molecule. As can be seen from Fig. 2, a successful fit to the experimental data was achieved with this approach. This success can be justified to some extent by the fact that a major portion of secondary electrons is ejected from the outermost MOs, with binding energies close to the ionization threshold.

The results of the best fits of Eq. (20) to the present experimental DDCS are depicted by the dashed curves in Fig. 2. Based on Pearson's Chi-square test, the semi-empirical formula with the best-fit parameter values was found to be consistent with the measured data within the confidence interval of 95%. As is evident from Fig. 2, the semi-empirical fits reproduce the measured data quite well at all primary and secondary electron energies within the experimental uncertainties.

The fit results were integrated over the solid angle to obtain the SDCS of Py between 4 eV and ($T$-I)/2. The SDCS at $E$ < 4 eV was determined by extrapolating the SDCS above 4 eV to lower energies. The extrapolation was performed by fitting a function based on the BEB model [33] to the SDCS:

$$g(w,t) = \sum_{k=1}^{3} z_k(t)[u_k(w+1) + u_k(t-w)], \qquad (23)$$

where $t = T/B$, $w = E/B$, $u_k(w+1) = (w+1)^{-k}$, $u_k(t-w) = (t-w)^{-k}$. The coefficients $z_k$ are fit parameters. The estimated uncertainty of the SDCS obtained in this way amounts to 28%. Figure 3 shows the SDCS obtained in this way in comparison to the values calculated using the BEB model [33] for six primary energies. The required binding and average electron kinetic energies in the MOs were calculated again using the Gausssian09 software [29] with the basis set 6-311++G. Unlike the SDCS determined from the experimental DDCS, the SDCS by the BEB model [33] was calculated for each MO separately and summed up afterwards.

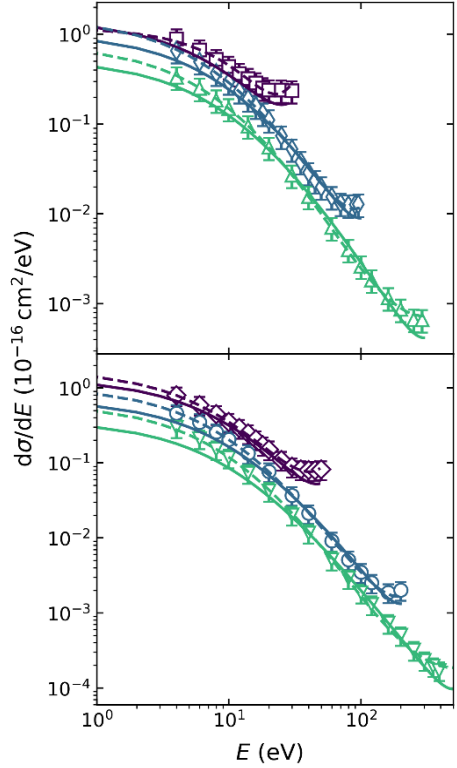

Fig. 3. SDCS of Py as a function of secondary electron energy *E*. To facilitate a clearer distinction of the results, the upper part of the figure depicts the SDCS for the primary energies 60 eV (□), 200 eV (◊) and 600 eV (△), while the bottom part represents those of the energies 100 eV (◇), 400 eV (○) and 1 keV (▽). The dashed curves were obtained by fitting Eq. (23) to the experimental SDCS. For comparison, the values calculated using the BEB model [33] are illustrated by solid curves.

The SDCS was further integrated over the secondary electron energy to obtain the TICS of Py. Figure 4 illustrates these TICS in comparison to the experimental data of Linert et al. [11], as well as to calculations with the BEB model [33] and the spherical complex optical potential model [13]. Although the experimental data of Linert et al. [11] agree with the results of this work within the experimental uncertainties, the former appear to trend lower than the latter. The BEB model reproduces the present results qualitatively well within the uncertainties, while our prior calculation [13] using the spherical complex optical potential model appears to slightly overestimate the TICS at electron energies below 100 eV. The latter describes inelastic scattering cross sections which also includes excitations. While ionization processes dominate the inelastic scattering at higher energies, these excitations are no longer negligible below 60 eV and can lead to an overestimation of the TICS below 60 eV.

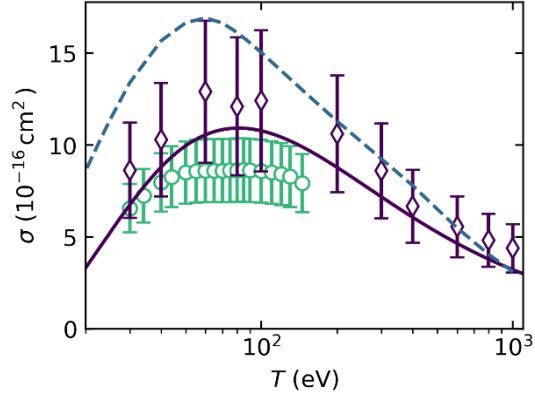

Fig. 4. Present TICS (◊) of Py as a function of primary electron energy $T$ in comparison to the experimental data (○) of Linert et al. [11], as well as to calculations using the BEB [33] (full line) and the spherical complex optical potential model [13] (dashed). The latter also includes rotational excitations, however, above $T > 60$ eV the dominant contribution to the inelastic scattering cross section comes from ionization processes.

## VII. Conclusion

A qualitatively good agreement was found between the DDCS of Py measured in this work and the calculations based on the DWBA. As expected, the DWBA calculation appears to better reproduce the experimental data at higher primary electron energies. Notably, a considerable difference between both data was observed at emission angles below 30°. In general, the emission of secondary electrons was nearly isotropic at low energies ($E \ll T$). As the energy of secondary electrons increases, binary collision peaks became more pronounced, reflecting a decrease in their width. The overestimate of the DDCS by the present theoretical approach at low emission angles is in part caused by the neglect of post collision interaction between the scattered and ejected electrons. A preliminary estimate suggests that post collision interaction may significantly reduce the DDCS at lower emission angles, particularly when the velocities of both electrons are similar. Proper consideration of post collision interaction could therefore lead to a better agreement between theory and experiment.

The experimental DDCS obtained in this work was well-fitted by a three-parameter semi-empirical formula which comprised two Lorentzian functions and a constant term. The SDCS obtained by integrating the fitted DDCS over the solid angle was satisfactorily reproduced by calculations using the BEB model within the experimental uncertainties. This is also the case for the TICS, which were obtained by integrating the SDCS over secondary electron energy.

**Acknowledgements**

The authors would like to thank Heike Nittmann for her assistance in conducting and evaluating the measurements. The technical assistance of Andreas Pausewang is also gratefully acknowledged.

**Appendix A1: Derivation of the computational form of the TDCS**

In the following, the computational form of Eq. (8) and subsequent derivations are explained for $T_{\text{dir}}$. The derivation of exchange terms can be performed in a similar way. Upon inserting Eqs. (11)-(14) into Eq. (8) and employing the Gaunt formula, we obtain

$$T_{\text{dir}} = \frac{\sqrt{2}/\pi^2}{k_0 k_A k_B} \sum_{lm} b_{lm}^p \sum_{L'L''M'} (-1)^{M'-m} \sum_{\lambda} \begin{pmatrix} l & \lambda & L'' \\ 0 & 0 & 0 \end{pmatrix} \begin{pmatrix} l & \lambda & L'' \\ m & -M' & M'-m \end{pmatrix}$$

$$\times \sum_{L} \begin{pmatrix} L' & \lambda & L \\ 0 & 0 & 0 \end{pmatrix} \begin{pmatrix} L' & \lambda & L \\ M' & -M' & 0 \end{pmatrix} R_{lmLL'L''\lambda}(k_0, k_A, k_B) \times Y_{L'M'}(\hat{\mathbf{k}}_A) Y_{L''M'-m}(\hat{\mathbf{k}}_B) \quad \text{(A1)}$$

with the radial integral

$$R_{lmLL'L''\lambda}(k_0, k_A, k_B) = i^{L-L'-L''} e^{i(\sigma_{L'}+\sigma_{L''})} \hat{l} \hat{L}^2 \hat{L}' \hat{L}'' \times$$

$$\int dr_A \int dr_B\, \chi_{L'}(k_A, r_A) \chi_{L''}(k_B, r_B) \frac{r_<^\lambda}{r_>^{\lambda+1}} u_{lm}(r_B) \chi_L(k_0, r_A)\,. \quad \text{(A2)}$$

The quantum numbers $L', M'$ and $L'', M''$ belong to the electrons with momentum $\mathbf{k}_A$ and $\mathbf{k}_B$, respectively, and the expressions in brackets are Wigner 3j symbols. In Eq. (A1), the z-axis is chosen as the electron incidence axis, meaning $\hat{\mathbf{k}}_0$ is parallel to the z-axis so that $M = 0$:

$$Y_{LM}^*(\hat{\mathbf{k}}_0) = Y_{L0}^*(\theta=0, \varphi) = \hat{L}/\sqrt{4\pi}. \quad \text{(A3)}$$

As explained in the main text, the molecular wavefunction (Eqs. (13)-(14)) is multiplied with the Wigner D-matrix according to Eq. (15) and the absolute square of the resulting **T**-matrix is averaged over different orientations. Reevaluating Eq. (A1) within the laboratory frame, the direct amplitude $T'_{\text{dir}}$ of a certain rotation $\alpha, \beta, \gamma$ reads

$$T'_{\text{dir}} = \frac{\sqrt{2}/\pi^2}{k_0 k_A k_B} \sum_{lm} b_{lm}^p \sum_{m_1=-l}^{l} D_{m_1,m}^l(\alpha, \beta, \gamma) S_{lmm_1}^{\text{dir}} \quad \text{(A4)}$$

with

$$S_{lmm_1}^{\text{dir}} = \sum_{L'L''M'} (-1)^{M'-m_1} \sum_{\lambda} \begin{pmatrix} l & \lambda & L'' \\ 0 & 0 & 0 \end{pmatrix} \begin{pmatrix} l & \lambda & L'' \\ m_1 & -M' & M'-m_1 \end{pmatrix} \quad \text{(A5)}$$

$$\times \sum_L \begin{pmatrix} L' & \lambda & L \\ 0 & 0 & 0 \end{pmatrix} \begin{pmatrix} L' & \lambda & L \\ M' & -M' & 0 \end{pmatrix} R_{lmLL'L''\lambda}(k_0, k_A, k_B)$$

$$\times Y_{L'M'}(\hat{\mathbf{k}}_A) Y_{L''M'-m_1}(\hat{\mathbf{k}}_B).$$

It follows for $|T'_{\text{dir}}|^2$:

$$|T'_{\text{dir}}|^2 = \frac{\sqrt{2}/\pi^2}{(k_0 k_A k_B)^2} \sum_{lm,l'm'} \sum_{m_1=-l}^{l} \sum_{m'_1=-l'}^{l'} b^p_{lm} b^p_{l'm'} \left( D^{l'}_{m'_1,m'} S^{\text{dir}}_{l'm'm'_1} \right)^* \left( D^{l}_{m_1,m} S^{\text{dir}}_{lmm_1} \right) \quad (A6)$$
$$+ c.c.$$

The averaging of $|T_{\text{dir}}|^2$ over different molecular orientations in the LF is given by

$$\overline{|T_{\text{dir}}|^2} = \frac{1}{8\pi^2} \int_0^{2\pi} d\alpha \int_0^{\pi} \sin\beta \, d\beta \int_0^{2\pi} d\gamma \, |T'_{\text{dir}}|^2. \quad (A7)$$

Utilizing the orthogonality relation of the Wigner D-matrices

$$\int\int\int d\alpha \, d\beta \, d\gamma \, D^{l'*}_{m'_1,m'}(\alpha,\beta,\gamma) \, D^{l}_{m_1,m}(\alpha,\beta,\gamma) = \frac{8\pi^2}{2l+1} \delta_{l'l} \delta_{m'_1 m_1} \delta_{m'm}, \quad (A8)$$

Eq. (A7) can be rewritten as

$$\overline{|T_{\text{dir}}|^2} = \frac{\sqrt{2}/\pi^2}{(k_0 k_A k_B)^2} \sum_{lm} \frac{(b^p_{lm})^2}{2l+1} \sum_{m_1=-l}^{l} |S^{\text{dir}}_{lmm_1}|^2. \quad (A9)$$

**Appendix A2: Derivation of the computational form of the DDCS**

The DDCS can be obtained in analytical form by integrating the orientation-averaged **T**-matrix over $\hat{\mathbf{k}}_A$

$$\frac{d^2\sigma_i}{d\Omega_B dE_B} = (2\pi)^4 \frac{n_i k_A k_B}{k_0} \int d\hat{\mathbf{k}}_A \, \overline{|T_{\text{dir}}|^2} + \text{exchange terms}. \quad (A10)$$

With $\overline{|T_{\text{dir}}|^2}$ given by Eq. (A9) and the orthogonality relation of spherical harmonics, the integration results in

$$\int d\hat{\mathbf{k}}_A \, \overline{|T_{\text{dir}}|^2} = \frac{\sqrt{2}/\pi^2}{(k_0 k_A k_B)^2} \sum_{lm} \frac{(b^p_{lm})^2}{2l+1} \sum_{m_1=-l}^{l} \int d\hat{\mathbf{k}}_A \, |S^{\text{dir}}_{lmm_1}|^2$$
$$= \sum_{lm} \frac{(b^p_{lm})^2}{2l+1} \sum_{m_1=-l}^{l} \sum_{L'M'} \left| \sum_{L''} Q_{L'M'L''lmm_1}(k_0, k_A, k_B) Y_{L''M'-m_1}(\hat{\mathbf{k}}_B) \right|^2, \quad (A11)$$

with

$$Q_{L'M'L''lmm_1}(k_0,k_A,k_B) = (-1)^{(M'-m_1)} \sum_\lambda \begin{pmatrix} l & \lambda & L'' \\ 0 & 0 & 0 \end{pmatrix} \begin{pmatrix} l & \lambda & L'' \\ m_1 & -M' & M'-m_1 \end{pmatrix}$$

$$\times \sum_L \begin{pmatrix} L' & \lambda & L \\ 0 & 0 & 0 \end{pmatrix} \begin{pmatrix} L' & \lambda & L \\ M' & -M' & 0 \end{pmatrix} R_{lmLL'L''\lambda}(k_0,k_A,k_B).$$

(A12)

**References**

[1] D. T. Goodhead, Int. J. Radiat. Biol. **65**, 7 (1994).

[2] M. A. Hill, Clin. Oncol. **32**, 75 (2020).

[3] H. Nikjoo, D. E. Charlton, and D. T. Goodhead, Adv. Space Res. **14**, 161 (1994).

[4] W. Friedland, M. Dingfelder, P. Kundrát, and P. Jacob, Mutat. Res. **711**, 28 (2011).

[5] C. Villagrasa, S. Meylan, G. Gonon, G. Gruel, U. Giesen, M. Bueno, and H. Rabus, EPJ Web Conf. **153**: 04019 (2017).

[6] R. D. White, M. J. Brunger, N. A. Garland, R. E. Robson, K. F. Ness, G. Garcia, J. de Urquijo, S. Dujko, and Z. L. Petrović, Eur. Phys. J. D **68:** 125 (2014).

[7] M. U. Bug, G. Hilgers, W. Y. Baek, and H. Rabus, Eur. Phys. J. D 68, 217 (2014).

[8] M. C. Fuss, A. G. Sanz, F. Blanco, J. C. Oller, P. Limao-Vieira, M. J. Brunger, and G. García, Phys. Rev. A **88**, 042702 (2013).

[9] J. B. Maljković, A. R. Milosavljević, F. Blanco, D. Šević, G. García, and B. P. Marinković, Phys. Rev. A **79**, 052706 (2009).

[10] P. Palihawadana, J. Sullivan, M. Brunger, C. Winstead, V. McKoy, G. Garcia, F. Blanco, and S. Buckman, Phys. Rev. A **84**, 062702 (2011).

[11] I. Linert, M. Dampc, B. Mielewska, and M. Zubeka, Eur. Phys. J. D **66**: 20 (2012).

[12] J. D. Builth-Williams, S. M. Bellm, D. B. Jones, Hari Chaluvadi, D. H. Madison, C. G. Ning, B. Lohmann, and M. J. Brunger, J. Chem. Phys. **136**, 024304 (2012).

[13] W. Y. Baek, A. Arndt, M. U. Bug, H. Rabus, and M. Wang, Phys. Rev. A. **88**, 032702 (2013).

[14] W. Y. Baek, M. U. Bug, and H. Rabus, Phys. Rev. A **89**, 062716 (2014).

[15] N. Sanna and F. A. Gianturco, Comput. Phys. Commun. **128,** 139 (2000).

[16] X. Xu, M. Gong, X. Li, S. B. Zhang, and X. Chen, J. Chem. Phys. **148**, 244104 (2018)

[17] G. B. da Silva, R. F. C. Neves, L. Chiari, D. B. Jones, E. Ali, D. H. Madison, C. G. Ning, K. L. Nixon, M. C. A. Lopes, and M. J. Brunger, J. Chem. Phys. **141**, 124307 (2014).

[18] A. Pandey and G. Purohit, Nucl. Inst. Meth. Phys. Res. B **547**, 165222 (2024).

[19] J. Gao, J. L. Peacher, and D. H. Madison, J. Chem. Phys. **123**, 204302 (2005).

[20] W. Y. Baek, M. Bug, H. Rabus, E. Gargioni, and B. Grosswendt, Phys. Rev. A **86**, 032702 (2012).

[21] W. Y. Baek, M. U. Bug, H. Nettelbeck, and H. Rabus, Eur. Phys. J. D **73**:61 (2019).

[22] R. T. Brinkmann and S. Trajmar, J. Phys. E **14**, 245 (1981).

[23] M.J. Brunger and S.J. Buckman, Phys. Rep. **357**, 215 (2002).

[24] S. K. Srivastava, A. Chutjian, and S. Trajmar, J. Chem. Phys. 63, 2659 (1975).
[25] A. E. Hughes and C. C. Phillips, Surf. Interface Anal. **4**, 5 (1982).
[26] I. E. McCarthy and Z. Xixiang, *Distorted-Wave Methods for Ionization*, Computational Atomic Physics, Ed. K. Bartschat, (Springer-Verlag, Berlin, Heidelberg 1996).
[27] N. Sanna, G. Morelli, S. Orlandini, M. Tacconi, and I. Baccarelli, Comput. Phys. Commun. **248**, 106970 (2020).
[28] P. G. Burke, N. Chandra, and F. A. Gianturco, J. Phys. B: Atom. Mol. Phys. **5**, 2212 (1972).
[29] GAUSSIAN 09, Gaussian, Inc., Wallingford, CT.
[30] International Organization for Standardization (ISO), *Guide to the Expression of uncertainty in Measurement* (ISO, Geneva, 1993).
[31] M.E. Rudd, Phys. Rev. A 44, 1644 (1991).
[32] S. J. Ward and J. H. Macek, Phys. Rev. A **49**, 1049 (1994).
[33] Y.-K. Kim and M. E. Rudd, Phys. Rev. A **50**, 3954 (1994).

**Supplemental Material**

TABLE I. Experimental results of this work. The DDCS is given in units of $10^{-18}$ cm$^2$/eVsr, while the SDCS and TICS are presented in units of $10^{-18}$ cm$^2$/eV and $10^{-18}$ cm$^2$, respectively. The numbers in the parentheses are the powers of ten by which the preceding number should be multiplied. The estimated overall uncertainties of the DDCS, SDCS, and TICS are 24%, 28%, and 30%, respectively.

| $T$=30 eV | | | | | DDCS | | | | | |
|---|---|---|---|---|---|---|---|---|---|---|
| $E$ (eV) | 25° | 35° | 45° | 60° | 75° | 90° | 105° | 120° | 135° | SDCS |
| 4 | 8.79 | 8.35 | 8.41 | 8.13 | 7.35 | 7.23 | 7.22 | 4.63 | 5.08 | 87.1 |
| 5 | 8.71 | 7.71 | 7.81 | 7.42 | 6.68 | 6.53 | 6.48 | 4.66 | 4.78 | 80.0 |
| 6 | 8.37 | 7.21 | 7.26 | 6.72 | 6.03 | 5.93 | 5.75 | 4.41 | 4.37 | 72.7 |
| 7 | 8.03 | 6.88 | 6.91 | 6.24 | 5.55 | 5.45 | 5.23 | 4.18 | 4.04 | 67.4 |
| 8 | 7.99 | 6.69 | 6.66 | 5.89 | 5.20 | 5.03 | 4.84 | 3.99 | 3.83 | 64.1 |
| 9 | 7.97 | 6.61 | 6.45 | 5.62 | 4.92 | 4.70 | 4.53 | 3.83 | 3.68 | 63.5 |
| 10 | 8.07 | 6.61 | 6.32 | 5.35 | 4.68 | 4.42 | 4.27 | 3.72 | 3.53 | 61.4 |
| TICS | | | | | | | | | | 8.62(+2) |
| $T$=40 eV | | | | | DDCS | | | | | |
| $E$ (eV) | 25° | 35° | 45° | 60° | 75° | 90° | 105° | 120° | 135° | SDCS |
| 4 | 8.22 | 7.35 | 7.57 | 7.54 | 7.46 | 7.38 | 6.85 | 6.50 | 6.82 | 90.1 |
| 5 | 7.44 | 6.59 | 6.77 | 6.73 | 6.63 | 6.58 | 6.16 | 5.81 | 5.99 | 80.0 |
| 6 | 6.75 | 5.97 | 6.08 | 5.98 | 5.86 | 5.76 | 5.41 | 5.15 | 5.17 | 71.1 |
| 7 | 6.43 | 5.55 | 5.57 | 5.42 | 5.27 | 5.09 | 4.81 | 4.57 | 4.56 | 63.2 |
| 8 | 6.11 | 5.26 | 5.19 | 4.97 | 4.81 | 4.58 | 4.33 | 4.14 | 4.11 | 58.6 |
| 9 | 5.61 | 5.02 | 4.90 | 4.60 | 4.41 | 4.16 | 3.97 | 3.81 | 3.75 | 53.8 |
| 10 | 5.31 | 4.88 | 4.67 | 4.29 | 4.08 | 3.81 | 3.68 | 3.53 | 3.45 | 51.0 |
| 12 | 5.48 | 4.81 | 4.38 | 3.84 | 3.56 | 3.32 | 3.22 | 3.10 | 3.08 | 46.3 |
| 14 | 5.87 | 4.91 | 4.26 | 3.55 | 3.23 | 2.94 | 2.87 | 2.80 | 2.80 | 43.5 |
| TICS | | | | | | | | | | 1.03(+3) |
| $T$=60 eV | | | | | DDCS | | | | | |
| $E$ (eV) | 25° | 35° | 45° | 60° | 75° | 90° | 105° | 120° | 135° | SDCS |
| 4 | 6.72 | 6.58 | 6.75 | 7.11 | 7.35 | 7.14 | 6.67 | 6.38 | 7.27 | 88.6 |
| 5 | 5.77 | 5.76 | 5.96 | 6.27 | 6.50 | 6.27 | 5.92 | 5.64 | 6.22 | 77.1 |
| 6 | 4.89 | 5.03 | 5.24 | 5.50 | 5.70 | 5.44 | 5.12 | 4.90 | 5.28 | 66.5 |
| 7  | 4.34 | 4.48 | 4.67 | 4.89 | 5.03 | 4.75 | 4.43 | 4.30 | 4.54 | 58.3 |
| 8  | 3.85 | 4.05 | 4.23 | 4.38 | 4.47 | 4.18 | 3.90 | 3.80 | 3.96 | 52.2 |
| 9  | 3.46 | 3.72 | 3.84 | 3.97 | 4.03 | 3.71 | 3.47 | 3.42 | 3.51 | 47.7 |
| 10 | 3.13 | 3.47 | 3.54 | 3.66 | 3.66 | 3.34 | 3.13 | 3.12 | 3.17 | 43.8 |
| 12 | 2.79 | 3.09 | 3.12 | 3.17 | 3.08 | 2.80 | 2.62 | 2.59 | 2.66 | 36.5 |
| 14 | 2.49 | 2.87 | 2.82 | 2.77 | 2.67 | 2.39 | 2.23 | 2.25 | 2.31 | 32.1 |
| 16 | 2.35 | 2.70 | 2.61 | 2.48 | 2.31 | 2.08 | 1.93 | 2.03 | 2.08 | 28.7 |
| 18 | 2.31 | 2.61 | 2.45 | 2.24 | 2.05 | 1.81 | 1.71 | 1.81 | 1.91 | 25.2 |
| 20 | 2.33 | 2.58 | 2.34 | 2.05 | 1.83 | 1.62 | 1.54 | 1.67 | 1.76 | 23.9 |
| 24 | 2.79 | 2.70 | 2.26 | 1.81 | 1.55 | 1.37 | 1.33 | 1.47 | 1.59 | 23.8 |
| TICS |   |   |   |   |   |   |   |   |   | 1.29(+3) |

| $T$=80 eV |  |  |  |  | DDCS |  |  |  |  |  |
|---|---|---|---|---|---|---|---|---|---|---|
| $E$ (eV) | 25° | 35° | 45° | 60° | 75° | 90° | 105° | 120° | 135° | SDCS |
| 4  | 6.43 | 5.97 | 6.21 | 6.41 | 6.33 | 6.47 | 6.18 | 6.26 | 6.07 | 78.4 |
| 5  | 5.73 | 5.18 | 5.44 | 5.62 | 5.52 | 5.61 | 5.28 | 5.35 | 5.22 | 68.1 |
| 6  | 4.56 | 4.52 | 4.72 | 4.87 | 4.79 | 4.78 | 4.49 | 4.50 | 4.40 | 58.6 |
| 7  | 3.68 | 3.94 | 4.18 | 4.27 | 4.21 | 4.11 | 3.86 | 3.83 | 3.74 | 50.4 |
| 8  | 3.29 | 3.50 | 3.75 | 3.82 | 3.74 | 3.63 | 3.36 | 3.31 | 3.25 | 45.1 |
| 9  | 3.09 | 3.16 | 3.36 | 3.45 | 3.34 | 3.25 | 2.96 | 2.90 | 2.86 | 39.8 |
| 10 | 2.74 | 2.86 | 3.08 | 3.14 | 3.03 | 2.92 | 2.65 | 2.56 | 2.54 | 36.2 |
| 12 | 2.07 | 2.47 | 2.62 | 2.67 | 2.54 | 2.36 | 2.14 | 2.08 | 2.06 | 29.1 |
| 14 | 1.85 | 2.17 | 2.29 | 2.28 | 2.15 | 2.00 | 1.80 | 1.74 | 1.76 | 25.6 |
| 16 | 1.61 | 1.91 | 2.01 | 1.98 | 1.82 | 1.67 | 1.50 | 1.47 | 1.52 | 21.6 |
| 18 | 1.50 | 1.72 | 1.77 | 1.71 | 1.56 | 1.40 | 1.27 | 1.25 | 1.30 | 18.5 |
| 20 | 1.36 | 1.56 | 1.60 | 1.50 | 1.33 | 1.20 | 1.09 | 1.09 | 1.14 | 16.2 |
| 24 | 1.28 | 1.43 | 1.39 | 1.22 | 1.04 | 0.93 | 0.86 | 0.88 | 0.93 | 13.5 |
| 28 | 1.38 | 1.42 | 1.29 | 1.06 | 0.87 | 0.77 | 0.72 | 0.76 | 0.84 | 12.5 |
| 32 | 1.67 | 1.51 | 1.29 | 0.98 | 0.76 | 0.67 | 0.64 | 0.71 | 0.79 | 11.9 |
| TICS |   |   |   |   |   |   |   |   |   | 1.21(+3) |

| $T$=100 eV |  |  |  |  | DDCS |  |  |  |  |  |
|---|---|---|---|---|---|---|---|---|---|---|
| $E$ (eV) | 25° | 35° | 45° | 60° | 75° | 90° | 105° | 120° | 135° | SDCS |
| 4  | 6.62 | 6.48 | 6.66 | 6.59 | 6.63 | 6.64 | 6.28 | 6.05 | 5.66 | 79.0 |
| 5  | 5.65 | 5.71 | 5.93 | 5.86 | 5.84 | 5.83 | 5.42 | 5.19 | 4.89 | 69.4 |
| 6  | 4.71 | 4.93 | 5.09 | 5.08 | 5.09 | 4.95 | 4.62 | 4.37 | 4.10 | 58.9 |
| 7  | 3.85 | 4.34 | 4.48 | 4.46 | 4.40 | 4.25 | 3.95 | 3.71 | 3.40 | 50.2 |
| 8  | 3.25 | 3.84 | 4.00 | 4.02 | 3.90 | 3.76 | 3.41 | 3.22 | 2.93 | 44.4 |
| 9  | 2.88 | 3.46 | 3.58 | 3.61 | 3.54 | 3.35 | 3.00 | 2.82 | 2.56 | 39.4 |
| 10 | 2.58 | 3.11 | 3.27 | 3.26 | 3.19 | 2.97 | 2.66 | 2.50 | 2.26 | 35.1 |
| 12 | 1.99 | 2.64 | 2.73 | 2.74 | 2.62 | 2.46 | 2.15 | 1.97 | 1.82 | 29.1 |
| 14 | 1.58 | 2.22 | 2.38 | 2.34 | 2.24 | 2.04 | 1.80 | 1.65 | 1.52 | 24.6 |
| 16 | 1.35 | 1.89 | 2.01 | 1.96 | 1.87 | 1.66 | 1.47 | 1.35 | 1.29 | 20.5 |
| 18 | 1.16 | 1.63 | 1.73 | 1.68 | 1.57 | 1.37 | 1.21 | 1.13 | 1.06 | 17.1 |
| 20 | 0.98 | 1.43 | 1.51 | 1.45 | 1.33 | 1.18 | 1.03 | 0.97 | 0.91 | 15.0 |
| 24 | 0.81 | 1.19 | 1.20 | 1.12 | 1.00 | 0.86 | 0.76 | 0.72 | 0.70 | 11.5 |
| 28 | 0.78 | 1.07 | 1.04 | 0.92 | 0.78 | 0.68 | 0.61 | 0.60 | 0.59 | 9.51 |
| 32 | 0.86 | 1.05 | 0.96 | 0.80 | 0.65 | 0.56 | 0.51 | 0.52 | 0.52 | 8.33 |
| 36 | 1.03 | 1.07 | 0.94 | 0.73 | 0.57 | 0.48 | 0.45 | 0.47 | 0.48 | 8.09 |
| 40 | 1.26 | 1.15 | 0.95 | 0.69 | 0.53 | 0.44 | 0.41 | 0.44 | 0.46 | 8.01 |
| 44 | 1.56 | 1.26 | 0.98 | 0.67 | 0.50 | 0.41 | 0.39 | 0.42 | 0.45 | 7.92 |
| TICS |   |   |   |   |   |   |   |   |   | 1.24(+3) |

| $T$=200 eV |  |  |  |  | DDCS |  |  |  |  |  |
|---|---|---|---|---|---|---|---|---|---|---|
| $E$ (eV) | 25° | 35° | 45° | 60° | 75° | 90° | 105° | 120° | 135° | SDCS |
| 4 | 5.99 | 5.47 | 5.37 | 5.27 | 5.28 | 5.54 | 5.30 | 5.02 | 4.83 | 65.8 |
| 5 | 5.00 | 4.81 | 4.75 | 4.71 | 4.67 | 4.87 | 4.55 | 4.32 | 4.15 | 57.6 |
| 6 | 4.20 | 4.19 | 4.20 | 4.14 | 4.10 | 4.20 | 3.85 | 3.67 | 3.50 | 49.2 |
| 7 | 3.62 | 3.67 | 3.71 | 3.63 | 3.60 | 3.63 | 3.32 | 3.10 | 2.93 | 42.1 |
| 8 | 3.12 | 3.27 | 3.27 | 3.22 | 3.19 | 3.19 | 2.92 | 2.67 | 2.52 | 36.8 |

| | | | | | | | | | |
|---|---|---|---|---|---|---|---|---|---|
| 9 | 2.66 | 2.92 | 2.93 | 2.90 | 2.88 | 2.85 | 2.57 | 2.34 | 2.20 | 32.6 |
| 10 | 2.29 | 2.62 | 2.63 | 2.64 | 2.62 | 2.58 | 2.27 | 2.05 | 1.91 | 29.3 |
| 12 | 1.77 | 2.17 | 2.19 | 2.24 | 2.17 | 2.09 | 1.85 | 1.63 | 1.51 | 23.5 |
| 14 | 1.37 | 1.75 | 1.80 | 1.86 | 1.80 | 1.68 | 1.48 | 1.30 | 1.26 | 19.1 |
| 16 | 1.09 | 1.42 | 1.48 | 1.54 | 1.49 | 1.38 | 1.19 | 1.04 | 0.96 | 15.8 |
| 18 | 0.88 | 1.16 | 1.24 | 1.30 | 1.24 | 1.14 | 0.98 | 0.85 | 0.77 | 13.2 |
| 20 | 0.66 | 0.97 | 1.05 | 1.10 | 1.06 | 0.95 | 0.81 | 0.70 | 0.64 | 11.1 |
| 25 | 0.43 | 0.65 | 0.71 | 0.76 | 0.71 | 0.63 | 0.52 | 0.46 | 0.43 | 7.35 |
| 30 | 0.29 | 0.46 | 0.51 | 0.55 | 0.51 | 0.43 | 0.36 | 0.32 | 0.30 | 5.26 |
| 35 | 0.25 | 0.35 | 0.39 | 0.41 | 0.37 | 0.30 | 0.25 | 0.23 | 0.22 | 3.80 |
| 40 | 0.20 | 0.28 | 0.32 | 0.33 | 0.28 | 0.22 | 0.18 | 0.17 | 0.17 | 2.87 |
| 45 | 0.20 | 0.24 | 0.27 | 0.27 | 0.22 | 0.17 | 0.14 | 0.13 | 0.13 | 2.34 |
| 50 | 0.21 | 0.22 | 0.24 | 0.23 | 0.18 | 0.13 | 0.11 | 0.10 | 0.10 | 1.99 |
| 60 | 0.23 | 0.20 | 0.21 | 0.18 | 0.13 | 8.9(-2) | 7.4(-2) | 7.1(-2) | 7.5(-2) | 1.55 |
| 70 | 0.28 | 0.21 | 0.20 | 0.15 | 9.9(-2) | 6.8(-2) | 5.9(-2) | 5.7(-2) | 6.1(-2) | 1.31 |
| 80 | 0.37 | 0.23 | 0.20 | 0.13 | 8.1(-2) | 6.0(-2) | 5.1(-2) | 5.1(-2) | 5.7(-2) | 1.25 |
| 90 | 0.51 | 0.27 | 0.20 | 0.12 | 7.4(-2) | 5.5(-2) | 4.7(-2) | 4.8(-2) | 5.2(-2) | 1.26 |
| TICS | | | | | | | | | | 1.06(+3) |

| $T$=300 eV | | | | | DDCS | | | | | |
|---|---|---|---|---|---|---|---|---|---|---|
| $E$ (eV) | 25° | 35° | 45° | 60° | 75° | 90° | 105° | 120° | 135° | SDCS |
| 4 | 4.72 | 4.37 | 4.32 | 4.36 | 4.23 | 4.30 | 4.13 | 3.93 | 3.89 | 52.5 |
| 5 | 4.16 | 3.89 | 3.86 | 3.90 | 3.76 | 3.82 | 3.61 | 3.42 | 3.35 | 46.2 |
| 6 | 3.58 | 3.38 | 3.35 | 3.41 | 3.28 | 3.31 | 3.08 | 2.90 | 2.81 | 39.6 |
| 7 | 3.08 | 2.92 | 2.90 | 2.98 | 2.87 | 2.88 | 2.63 | 2.43 | 2.35 | 34.0 |
| 8 | 2.68 | 2.58 | 2.56 | 2.63 | 2.55 | 2.53 | 2.29 | 2.09 | 2.00 | 29.6 |
| 9 | 2.33 | 2.32 | 2.29 | 2.37 | 2.29 | 2.25 | 2.01 | 1.83 | 1.73 | 26.2 |
| 10 | 2.05 | 2.08 | 2.07 | 2.17 | 2.08 | 2.03 | 1.80 | 1.62 | 1.52 | 23.5 |
| 12 | 1.57 | 1.63 | 1.72 | 1.78 | 1.75 | 1.67 | 1.49 | 1.33 | 1.25 | 19.3 |
| 14 | 1.27 | 1.38 | 1.42 | 1.48 | 1.43 | 1.37 | 1.20 | 1.03 | 0.97 | 15.6 |
| 16 | 1.06 | 1.15 | 1.18 | 1.26 | 1.19 | 1.15 | 0.97 | 0.83 | 0.77 | 12.9 |
| 18 | 0.85 | 0.95 | 0.98 | 1.06 | 1.01 | 0.95 | 0.80 | 0.68 | 0.63 | 10.7 |
| 20 | 0.69 | 0.78 | 0.82 | 0.91 | 0.87 | 0.80 | 0.66 | 0.57 | 0.52 | 8.97 |
| 25 | 0.43 | 0.52 | 0.56 | 0.63 | 0.60 | 0.54 | 0.44 | 0.37 | 0.34 | 6.03 |
| 30 | 0.29 | 0.36 | 0.40 | 0.46 | 0.44 | 0.38 | 0.30 | 0.26 | 0.24 | 4.31 |
| 35 | 0.21 | 0.27 | 0.30 | 0.36 | 0.33 | 0.28 | 0.22 | 0.19 | 0.18 | 3.20 |
| 40 | 0.16 | 0.21 | 0.24 | 0.28 | 0.26 | 0.21 | 0.16 | 0.14 | 0.14 | 2.46 |
| 45 | 0.13 | 0.17 | 0.20 | 0.23 | 0.20 | 0.16 | 0.12 | 0.10 | 0.10 | 1.93 |
| 50 | 0.10 | 0.14 | 0.17 | 0.20 | 0.16 | 0.12 | 9.1(-2) | 7.9(-2) | 7.8(-2) | 1.55 |
| 60 | 7.6(-2) | 0.10 | 0.13 | 0.15 | 0.11 | 7.7(-2) | 5.8(-2) | 4.7(-2) | 5.0(-2) | 1.07 |
| 70 | 6.0(-2) | 8.6(-2) | 0.11 | 0.12 | 8.1(-2) | 5.3(-2) | 3.8(-2) | 3.4(-2) | 3.6(-2) | 0.81 |
| 80 | 5.3(-2) | 7.8(-2) | 9.4(-2) | 9.5(-2) | 5.9(-2) | 3.9(-2) | 2.7(-2) | 2.4(-2) | 2.6(-2) | 0.64 |
| 90 | 5.7(-2) | 7.4(-2) | 8.6(-2) | 8.0(-2) | 4.6(-2) | 2.8(-2) | 2.0(-2) | 1.8(-2) | 2.0(-2) | 0.54 |
| 100 | 5.6(-2) | 7.5(-2) | 8.3(-2) | 7.0(-2) | 3.7(-2) | 2.2(-2) | 1.7(-2) | 1.5(-2) | 1.6(-2) | 0.48 |
| 120 | 7.1(-2) | 8.7(-2) | 8.3(-2) | 5.3(-2) | 2.5(-2) | 1.6(-2) | 1.2(-2) | 1.1(-2) | 1.2(-2) | 0.43 |
| 140 | 0.11 | 0.11 | 8.5(-2) | 4.4(-2) | 2.2(-2) | 1.5(-2) | 1.1(-2) | 1.0(-2) | 1.1(-2) | 0.44 |
| TICS | | | | | | | | | | 8.6(+2) |

| $T$=400 eV | | | | | DDCS | | | | | |
|---|---|---|---|---|---|---|---|---|---|---|
| $E$ (eV) | 25° | 35° | 45° | 60° | 75° | 90° | 105° | 120° | 135° | SDCS |
| 4 | 3.83 | 4.04 | 3.76 | 3.83 | 3.76 | 3.63 | 3.52 | 3.40 | 3.49 | 45.8 |
| 5 | 3.41 | 3.63 | 3.38 | 3.44 | 3.36 | 3.23 | 3.10 | 2.96 | 3.02 | 40.5 |
| 6 | 2.90 | 3.16 | 2.95 | 3.01 | 2.92 | 2.79 | 2.66 | 2.50 | 2.54 | 34.7 |
| 7 | 2.46 | 2.70 | 2.55 | 2.62 | 2.52 | 2.41 | 2.27 | 2.10 | 2.12 | 29.6 |
| 8 | 2.10 | 2.36 | 2.23 | 2.31 | 2.22 | 2.12 | 1.97 | 1.80 | 1.79 | 25.6 |
| 9 | 1.82 | 2.13 | 1.98 | 2.10 | 1.99 | 1.89 | 1.73 | 1.56 | 1.54 | 22.7 |
| 10 | 1.60 | 1.93 | 1.79 | 1.91 | 1.82 | 1.72 | 1.55 | 1.39 | 1.35 | 20.4 |
| 12 | 1.28 | 1.54 | 1.48 | 1.55 | 1.51 | 1.39 | 1.27 | 1.09 | 1.13 | 16.6 |
| 14 | 1.04 | 1.25 | 1.21 | 1.28 | 1.25 | 1.15 | 1.02 | 0.86 | 0.86 | 13.4 |
| 16 | 0.83 | 1.03 | 0.99 | 1.08 | 1.05 | 0.97 | 0.83 | 0.70 | 0.67 | 11.0 |

| E (eV) | 25° | 35° | 45° | 60° | 75° | 90° | 105° | 120° | 135° | SDCS |
|---|---|---|---|---|---|---|---|---|---|---|
| 18 | 0.66 | 0.84 | 0.84 | 0.92 | 0.89 | 0.81 | 0.69 | 0.57 | 0.55 | 9.13 |
| 20 | 0.54 | 0.70 | 0.70 | 0.78 | 0.76 | 0.69 | 0.58 | 0.47 | 0.45 | 7.67 |
| 25 | 0.34 | 0.46 | 0.47 | 0.54 | 0.53 | 0.47 | 0.38 | 0.31 | 0.29 | 5.16 |
| 30 | 0.22 | 0.32 | 0.33 | 0.40 | 0.39 | 0.34 | 0.27 | 0.22 | 0.21 | 3.70 |
| 35 | 0.16 | 0.24 | 0.25 | 0.31 | 0.30 | 0.25 | 0.20 | 0.16 | 0.15 | 2.77 |
| 40 | 0.12 | 0.18 | 0.20 | 0.25 | 0.23 | 0.19 | 0.14 | 0.12 | 0.11 | 2.12 |
| 45 | 9.8(-2) | 0.14 | 0.16 | 0.20 | 0.19 | 0.14 | 0.11 | 8.7(-2) | 8.6(-2) | 1.67 |
| 50 | 7.5(-2) | 0.11 | 0.13 | 0.17 | 0.16 | 0.11 | 8.0(-2) | 6.7(-2) | 6.5(-2) | 1.33 |
| 60 | 4.9(-2) | 8.0(-2) | 9.6(-2) | 0.13 | 0.11 | 7.2(-2) | 5.1(-2) | 4.2(-2) | 4.3(-2) | 0.91 |
| 70 | 3.8(-2) | 6.2(-2) | 7.8(-2) | 0.10 | 7.8(-2) | 4.8(-2) | 3.5(-2) | 2.8(-2) | 2.9(-2) | 0.67 |
| 80 | 3.1(-2) | 5.2(-2) | 6.3(-2) | 8.2(-2) | 6.0(-2) | 3.4(-2) | 2.4(-2) | 2.0(-2) | 2.1(-2) | 0.51 |
| 90 | 2.9(-2) | 4.5(-2) | 5.6(-2) | 7.0(-2) | 4.5(-2) | 2.5(-2) | 1.7(-2) | 1.5(-2) | 1.6(-2) | 0.42 |
| 100 | 2.4(-2) | 4.1(-2) | 5.3(-2) | 6.1(-2) | 3.5(-2) | 1.9(-2) | 1.3(-2) | 1.3(-2) | 1.2(-2) | 0.35 |
| 120 | 2.1(-2) | 3.7(-2) | 4.5(-2) | 4.5(-2) | 2.2(-2) | 1.2(-2) | 8.0(-3) | 7.0(-3) | 7.3(-3) | 0.25 |
| 140 | 2.2(-2) | 3.7(-2) | 4.3(-2) | 3.5(-2) | 1.4(-2) | 7.9(-3) | 5.4(-3) | 4.7(-3) | 5.0(-3) | 0.21 |
| 160 | 2.6(-2) | 4.2(-2) | 4.3(-2) | 2.7(-2) | 1.1(-2) | 6.1(-3) | 4.4(-3) | 3.5(-3) | 3.7(-3) | 0.19 |
| 180 | 3.6(-2) | 5.1(-2) | 4.3(-2) | 2.1(-2) | 9.0(-3) | 5.4(-3) | 4.0(-3) | 3.2(-3) | 3.3(-3) | 0.19 |
| 190 | 4.3(-2) | 5.6(-2) | 4.5(-2) | 2.0(-2) | 7.8(-3) | 5.0(-3) | 3.7(-3) | 3.2(-3) | 3.5(-3) | 0.19 |
| TICS | | | | | | | | | | 6.66(+2) |

| $T$=600 eV | | | | | DDCS | | | | | |
|---|---|---|---|---|---|---|---|---|---|---|
| $E$ (eV) | 25° | 35° | 45° | 60° | 75° | 90° | 105° | 120° | 135° | SDCS |
| 4 | 2.60 | 2.78 | 2.82 | 2.96 | 3.01 | 3.23 | 2.78 | 2.20 | 2.26 | 33.2 |
| 5 | 2.19 | 2.48 | 2.51 | 2.65 | 2.68 | 2.87 | 2.45 | 1.92 | 1.97 | 29.3 |
| 6 | 1.80 | 2.14 | 2.17 | 2.28 | 2.31 | 2.46 | 2.09 | 1.63 | 1.68 | 25.0 |
| 7 | 1.52 | 1.82 | 1.84 | 1.95 | 1.99 | 2.09 | 1.78 | 1.36 | 1.44 | 21.3 |
| 8 | 1.28 | 1.57 | 1.61 | 1.71 | 1.74 | 1.82 | 1.53 | 1.16 | 1.25 | 18.5 |
| 9 | 1.12 | 1.40 | 1.45 | 1.54 | 1.56 | 1.63 | 1.36 | 1.02 | 1.08 | 16.4 |
| 10 | 0.99 | 1.26 | 1.33 | 1.42 | 1.43 | 1.47 | 1.23 | 0.92 | 0.94 | 14.8 |
| 12 | 0.76 | 1.01 | 1.05 | 1.14 | 1.16 | 1.20 | 0.96 | 0.72 | 0.72 | 11.7 |
| 14 | 0.60 | 0.82 | 0.85 | 0.95 | 0.96 | 0.99 | 0.79 | 0.57 | 0.55 | 9.50 |
| 16 | 0.47 | 0.67 | 0.71 | 0.79 | 0.80 | 0.82 | 0.65 | 0.46 | 0.44 | 7.81 |
| 18 | 0.38 | 0.56 | 0.59 | 0.67 | 0.69 | 0.69 | 0.54 | 0.38 | 0.36 | 6.52 |
| 20 | 0.29 | 0.46 | 0.50 | 0.57 | 0.59 | 0.59 | 0.45 | 0.31 | 0.30 | 5.49 |
| 25 | 0.20 | 0.30 | 0.33 | 0.40 | 0.41 | 0.41 | 0.30 | 0.21 | 0.19 | 3.73 |
| 30 | 0.13 | 0.21 | 0.23 | 0.29 | 0.31 | 0.30 | 0.22 | 0.14 | 0.13 | 2.69 |
| 35 | 9.7(-2) | 0.15 | 0.18 | 0.23 | 0.24 | 0.23 | 0.16 | 0.11 | 0.10 | 2.03 |
| 40 | 7.2(-2) | 0.11 | 0.14 | 0.18 | 0.19 | 0.17 | 0.12 | 7.6(-2) | 7.4(-2) | 1.56 |
| 45 | 6.3(-2) | 9.0(-2) | 0.11 | 0.15 | 0.15 | 0.14 | 8.8(-2) | 5.8(-2) | 5.6(-2) | 1.24 |
| 50 | 4.8(-2) | 7.1(-2) | 8.8(-2) | 0.13 | 0.13 | 0.11 | 6.8(-2) | 4.4(-2) | 4.3(-2) | 0.99 |
| 60 | 3.7(-2) | 4.9(-2) | 6.5(-2) | 9.5(-2) | 9.3(-2) | 7.3(-2) | 4.4(-2) | 2.8(-2) | 2.9(-2) | 0.70 |
| 70 | 3.0(-2) | 3.7(-2) | 5.0(-2) | 7.6(-2) | 7.2(-2) | 5.0(-2) | 2.9(-2) | 1.9(-2) | 1.9(-2) | 0.52 |
| 80 | 2.3(-2) | 2.9(-2) | 4.0(-2) | 6.3(-2) | 5.6(-2) | 3.7(-2) | 2.0(-2) | 1.3(-2) | 1.4(-2) | 0.40 |
| 90 | 2.1(-2) | 2.4(-2) | 3.4(-2) | 5.3(-2) | 4.5(-2) | 2.7(-2) | 1.5(-2) | 1.0(-2) | 1.1(-2) | 0.32 |
| 100 | 1.9(-2) | 2.0(-2) | 2.9(-2) | 4.7(-2) | 3.6(-2) | 2.0(-2) | 1.1(-2) | 7.3(-3) | 8.0(-3) | 0.26 |
| 120 | 1.5(-2) | 1.6(-2) | 2.4(-2) | 3.6(-2) | 2.4(-2) | 1.2(-2) | 6.4(-3) | 4.2(-3) | 4.8(-3) | 0.18 |
| 140 | 1.4(-2) | 1.3(-2) | 2.1(-2) | 3.0(-2) | 1.6(-2) | 7.6(-3) | 4.3(-3) | 2.9(-3) | 3.3(-3) | 0.14 |
| 160 | 1.3(-2) | 1.2(-2) | 1.9(-2) | 2.5(-2) | 1.1(-2) | 5.4(-3) | 3.1(-3) | 2.2(-3) | 2.5(-3) | 0.12 |
| 180 | 1.3(-2) | 1.2(-2) | 1.9(-2) | 2.0(-2) | 8.1(-3) | 4.0(-3) | 2.4(-3) | 1.8(-3) | 2.1(-3) | 9.8(-2) |
| 200 | 1.5(-2) | 1.2(-2) | 1.8(-2) | 1.7(-2) | 6.0(-3) | 3.2(-3) | 2.0(-3) | 1.4(-3) | 1.7(-3) | 8.7(-2) |
| 250 | 1.4(-2) | 1.6(-2) | 1.8(-2) | 9.4(-3) | 2.3(-3) | 1.3(-3) | 1.0(-3) | 6.5(-4) | 8.2(-4) | 6.7(-2) |
| TICS | | | | | | | | | | 5.55(+2) |

| $T$=800 eV | | | | | DDCS | | | | | |
|---|---|---|---|---|---|---|---|---|---|---|
| $E$ (eV) | 25° | 35° | 45° | 60° | 75° | 90° | 105° | 120° | 135° | SDCS |
| 4 | 2.87 | 2.99 | 2.40 | 2.35 | 2.43 | 2.43 | 2.14 | 2.20 | 2.24 | 29.9 |
| 5 | 2.48 | 2.56 | 2.13 | 2.07 | 2.15 | 2.14 | 1.90 | 1.93 | 1.96 | 26.3 |
| 6 | 2.09 | 2.15 | 1.86 | 1.79 | 1.86 | 1.84 | 1.63 | 1.63 | 1.65 | 22.4 |
| 7 | 1.76 | 1.80 | 1.61 | 1.53 | 1.58 | 1.56 | 1.38 | 1.38 | 1.37 | 18.9 |
| 8 | 1.52 | 1.52 | 1.43 | 1.35 | 1.37 | 1.35 | 1.19 | 1.20 | 1.16 | 16.4 |

| E (eV) | 25° | 35° | 45° | 60° | 75° | 90° | 105° | 120° | 135° | SDCS |
|---|---|---|---|---|---|---|---|---|---|---|
| 9 | 1.32 | 1.31 | 1.24 | 1.21 | 1.22 | 1.19 | 1.06 | 1.06 | 1.01 | 14.4 |
| 10 | 1.15 | 1.15 | 1.05 | 1.07 | 1.10 | 1.08 | 0.95 | 0.91 | 0.95 | 12.9 |
| 12 | 0.87 | 0.90 | 0.85 | 0.86 | 0.89 | 0.87 | 0.75 | 0.72 | 0.73 | 10.2 |
| 14 | 0.67 | 0.71 | 0.70 | 0.71 | 0.74 | 0.71 | 0.61 | 0.57 | 0.55 | 8.10 |
| 16 | 0.52 | 0.57 | 0.58 | 0.59 | 0.62 | 0.59 | 0.50 | 0.46 | 0.43 | 6.61 |
| 18 | 0.42 | 0.47 | 0.47 | 0.50 | 0.53 | 0.50 | 0.41 | 0.38 | 0.35 | 5.49 |
| 20 | 0.34 | 0.37 | 0.40 | 0.42 | 0.45 | 0.42 | 0.35 | 0.31 | 0.29 | 4.59 |
| 25 | 0.21 | 0.24 | 0.26 | 0.29 | 0.32 | 0.29 | 0.23 | 0.21 | 0.19 | 3.08 |
| 30 | 0.14 | 0.16 | 0.19 | 0.22 | 0.24 | 0.22 | 0.17 | 0.14 | 0.13 | 2.22 |
| 35 | 0.10 | 0.12 | 0.14 | 0.16 | 0.18 | 0.17 | 0.12 | 0.11 | 9.7(-2) | 1.66 |
| 40 | 7.5(-2) | 8.9(-2) | 0.11 | 0.13 | 0.14 | 0.13 | 9.1(-2) | 7.8(-2) | 7.3(-2) | 1.27 |
| 45 | 5.7(-2) | 7.0(-2) | 8.3(-2) | 0.11 | 0.11 | 0.10 | 7.0(-2) | 5.9(-2) | 5.3(-2) | 1.00 |
| 50 | 4.4(-2) | 5.6(-2) | 6.7(-2) | 8.9(-2) | 9.7(-2) | 8.2(-2) | 5.5(-2) | 4.5(-2) | 4.2(-2) | 0.81 |
| 60 | 3.0(-2) | 3.6(-2) | 4.8(-2) | 6.6(-2) | 7.1(-2) | 5.6(-2) | 3.4(-2) | 2.8(-2) | 2.7(-2) | 0.56 |
| 70 | 2.1(-2) | 2.7(-2) | 3.6(-2) | 5.3(-2) | 5.6(-2) | 4.0(-2) | 2.4(-2) | 2.0(-2) | 1.9(-2) | 0.40 |
| 80 | 1.6(-2) | 2.1(-2) | 2.8(-2) | 4.3(-2) | 4.5(-2) | 2.9(-2) | 1.7(-2) | 1.4(-2) | 1.3(-2) | 0.32 |
| 90 | 1.2(-2) | 1.6(-2) | 2.3(-2) | 3.7(-2) | 3.6(-2) | 2.2(-2) | 1.2(-2) | 9.9(-3) | 9.8(-3) | 0.25 |
| 100 | 1.0(-2) | 1.4(-2) | 1.9(-2) | 3.2(-2) | 3.0(-2) | 1.7(-2) | 9.0(-3) | 7.2(-3) | 7.3(-3) | 0.20 |
| 120 | 7.5(-3) | 9.7(-3) | 1.5(-2) | 2.6(-2) | 2.2(-2) | 1.0(-2) | 5.5(-3) | 4.1(-3) | 4.5(-3) | 0.14 |
| 140 | 5.9(-3) | 7.9(-3) | 1.2(-2) | 2.2(-2) | 1.5(-2) | 6.5(-3) | 3.7(-3) | 3.1(-3) | 3.0(-3) | 0.11 |
| 160 | 5.1(-3) | 6.6(-3) | 1.1(-2) | 1.8(-2) | 1.1(-2) | 4.7(-3) | 2.4(-3) | 2.1(-3) | 2.3(-3) | 8.5(-2) |
| 180 | 4.8(-3) | 5.9(-3) | 9.7(-3) | 1.6(-2) | 8.5(-3) | 3.3(-3) | 1.9(-3) | 1.5(-3) | 1.8(-3) | 7.1(-2) |
| 200 | 4.5(-3) | 5.4(-3) | 9.2(-3) | 1.4(-2) | 6.1(-3) | 2.6(-3) | 1.6(-3) | 1.3(-3) | 1.6(-3) | 6.0(-2) |
| 250 | 4.0(-3) | 4.5(-3) | 8.7(-3) | 9.1(-3) | 2.9(-3) | 1.3(-3) | 8.9(-4) | 7.9(-4) | 9.1(-4) | 4.0(-2) |
| 300 | 4.3(-3) | 4.4(-3) | 9.2(-3) | 5.8(-3) | 1.6(-3) | 8.4(-4) | 5.7(-4) | 5.1(-4) | 5.8(-4) | 3.2(-2) |
| 350 | 5.5(-3) | 5.6(-3) | 9.8(-3) | 3.7(-3) | 1.1(-3) | 6.1(-4) | 4.3(-4) | 3.8(-4) | 4.1(-4) | 2.9(-2) |
| TICS | | | | | | | | | | 4.81(+2) |

| T=1 keV | | | | | DDCS | | | | | |
|---|---|---|---|---|---|---|---|---|---|---|
| E (eV) | 25° | 35° | 45° | 60° | 75° | 90° | 105° | 120° | 135° | SDCS |
| 4 | 2.15 | 2.31 | 2.28 | 2.87 | 2.53 | 2.98 | 2.51 | 1.79 | 2.31 | 27.6 |
| 5 | 1.70 | 1.95 | 1.96 | 2.39 | 2.20 | 2.48 | 2.08 | 1.51 | 1.96 | 23.5 |
| 6 | 1.38 | 1.65 | 1.69 | 1.99 | 1.84 | 2.10 | 1.72 | 1.26 | 1.63 | 20.1 |
| 7 | 1.09 | 1.41 | 1.45 | 1.71 | 1.52 | 1.77 | 1.41 | 1.05 | 1.35 | 17.4 |
| 8 | 0.90 | 1.23 | 1.32 | 1.53 | 1.34 | 1.51 | 1.18 | 0.90 | 1.12 | 15.2 |
| 9 | 0.80 | 1.06 | 1.16 | 1.33 | 1.22 | 1.32 | 1.02 | 0.79 | 0.99 | 13.3 |
| 10 | 0.73 | 0.91 | 0.95 | 1.13 | 1.08 | 1.17 | 0.90 | 0.69 | 0.90 | 11.8 |
| 12 | 0.56 | 0.72 | 0.74 | 0.91 | 0.85 | 0.94 | 0.70 | 0.52 | 0.66 | 9.37 |
| 14 | 0.50 | 0.57 | 0.61 | 0.75 | 0.69 | 0.77 | 0.55 | 0.41 | 0.50 | 7.58 |
| 16 | 0.43 | 0.46 | 0.49 | 0.62 | 0.57 | 0.63 | 0.46 | 0.34 | 0.40 | 6.23 |
| 18 | 0.34 | 0.38 | 0.41 | 0.51 | 0.48 | 0.53 | 0.39 | 0.27 | 0.32 | 5.19 |
| 20 | 0.28 | 0.32 | 0.34 | 0.44 | 0.42 | 0.46 | 0.33 | 0.23 | 0.27 | 4.38 |
| 25 | 0.19 | 0.20 | 0.23 | 0.30 | 0.29 | 0.31 | 0.23 | 0.15 | 0.17 | 2.98 |
| 30 | 0.11 | 0.14 | 0.16 | 0.22 | 0.21 | 0.23 | 0.17 | 0.10 | 0.12 | 2.13 |
| 35 | 8.4(-2) | 9.9(-2) | 0.12 | 0.17 | 0.16 | 0.17 | 0.12 | 7.6(-2) | 8.5(-2) | 1.58 |
| 40 | 6.2(-2) | 7.4(-2) | 9.0(-2) | 0.13 | 0.13 | 0.14 | 9.1(-2) | 5.6(-2) | 6.3(-2) | 1.21 |
| 45 | 4.1(-2) | 5.7(-2) | 6.9(-2) | 0.10 | 0.11 | 0.10 | 6.9(-2) | 4.2(-2) | 4.8(-2) | 0.95 |
| 50 | 3.6(-2) | 4.4(-2) | 5.6(-2) | 8.7(-2) | 8.9(-2) | 8.4(-2) | 5.6(-2) | 3.2(-2) | 3.7(-2) | 0.76 |
| 60 | 2.4(-2) | 3.0(-2) | 3.9(-2) | 6.4(-2) | 6.8(-2) | 6.0(-2) | 3.5(-2) | 2.0(-2) | 2.3(-2) | 0.51 |
| 70 | 1.8(-2) | 2.1(-2) | 2.8(-2) | 5.0(-2) | 5.3(-2) | 4.4(-2) | 2.4(-2) | 1.3(-2) | 1.6(-2) | 0.37 |
| 80 | 1.2(-2) | 1.6(-2) | 2.2(-2) | 4.1(-2) | 4.3(-2) | 3.3(-2) | 1.6(-2) | 9.6(-3) | 1.2(-2) | 0.27 |
| 90 | 8.8(-3) | 1.2(-2) | 1.8(-2) | 3.5(-2) | 3.4(-2) | 2.5(-2) | 1.2(-2) | 7.1(-3) | 8.7(-3) | 0.21 |
| 100 | 7.4(-3) | 1.0(-2) | 1.6(-2) | 3.0(-2) | 2.9(-2) | 2.0(-2) | 9.0(-3) | 5.2(-3) | 6.6(-3) | 0.17 |
| 120 | 5.7(-3) | 7.2(-3) | 1.1(-2) | 2.3(-2) | 2.1(-2) | 1.2(-2) | 5.2(-3) | 3.3(-3) | 4.0(-3) | 0.11 |
| 140 | 3.4(-3) | 5.5(-3) | 8.8(-3) | 1.9(-2) | 1.5(-2) | 7.4(-3) | 3.5(-3) | 2.2(-3) | 2.7(-3) | 8.2(-2) |
| 160 | 3.4(-3) | 4.3(-3) | 7.3(-3) | 1.6(-2) | 1.1(-2) | 5.1(-3) | 2.4(-3) | 1.5(-3) | 2.0(-3) | 6.3(-2) |
| 180 | 2.8(-3) | 3.7(-3) | 6.4(-3) | 1.4(-2) | 8.2(-3) | 3.6(-3) | 1.8(-3) | 1.2(-3) | 1.5(-3) | 5.1(-2) |
| 200 | 2.7(-3) | 3.3(-3) | 5.8(-3) | 1.2(-2) | 6.2(-3) | 2.7(-3) | 1.4(-3) | 9.8(-4) | 1.4(-3) | 4.2(-2) |
| 250 | 2.1(-3) | 2.5(-3) | 4.8(-3) | 8.7(-3) | 3.0(-3) | 1.3(-3) | 6.8(-4) | 5.9(-4) | 7.4(-4) | 3.2(-2) |
| 300 | 2.6(-3) | 2.4(-3) | 4.5(-3) | 6.2(-3) | 1.7(-3) | 8.1(-4) | 4.3(-4) | 3.2(-4) | 4.1(-4) | 2.3(-2) |

| 350 | 2.3(-3) | 2.6(-3) | 5.2(-3) | 4.3(-3) | 1.0(-3) | 4.9(-4) | 3.3(-4) | 2.1(-4) | 2.8(-4) | 1.9(-2) |
| 400 | 2.6(-3) | 2.9(-3) | 5.5(-3) | 2.5(-3) | 6.1(-4) | 3.3(-4) | 2.3(-4) | 1.6(-4) | 1.9(-4) | 1.7(-2) |
| 450 | 3.5(-3) | 3.9(-3) | 5.8(-3) | 1.5(-3) | 4.9(-4) | 3.6(-4) | 1.6(-4) | 1.4(-4) | 2.4(-4) | 1.7(-2) |
| TICS | | | | | | | | | | 4.38(+2) |